\documentclass[aps,twocolumn,superscriptaddress,shortbibliography]{revtex4-1}

\usepackage{amsmath,amssymb}
\usepackage{graphicx}
\usepackage{hyperref}
\usepackage{braket}
\usepackage{xcolor}

\begin{document}
\title{All Optical Neural Network with Nonlinear Activation Functions }
\author{Ying Zuo}\thanks{These authors contributed equally to this work.}
\affiliation{Department of Physics, The Hong Kong University of Science and Technology, Clear Water Bay, Kowloon, Hong Kong, China}

\author{Bohan Li}\thanks{These authors contributed equally to this work.}
\affiliation{Department of Physics, The Hong Kong University of Science and Technology, Clear Water Bay, Kowloon, Hong Kong, China}

\author{Yujun Zhao}\thanks{These authors contributed equally to this work.}
\affiliation{Department of Physics, The Hong Kong University of Science and Technology, Clear Water Bay, Kowloon, Hong Kong, China}

\author{Yue Jiang}
\affiliation{Department of Physics, The Hong Kong University of Science and Technology, Clear Water Bay, Kowloon, Hong Kong, China}

\author{You-Chiuan Chen}
\affiliation{Department of Physics, The Hong Kong University of Science and Technology, Clear Water Bay, Kowloon, Hong Kong, China}

\author{Peng Chen}
\affiliation{Department of Physics, The Hong Kong University of Science and Technology, Clear Water Bay, Kowloon, Hong Kong, China}

\author{Gyu-Boong Jo}
\affiliation{Department of Physics, The Hong Kong University of Science and Technology, Clear Water Bay, Kowloon, Hong Kong, China}

\author{Junwei Liu}
\email{liuj@ust.hk}
\affiliation{Department of Physics, The Hong Kong University of Science and Technology, Clear Water Bay, Kowloon, Hong Kong, China}

\author{Shengwang Du}
\email{dusw@ust.hk}
\affiliation{Department of Physics, The Hong Kong University of Science and Technology, Clear Water Bay, Kowloon, Hong Kong, China}

\begin{abstract}
\noindent \textbf{Abstract}: Artificial neural networks (ANNs) have now been widely used for industry applications and also played more important roles in fundamental researches. Although most ANN hardware systems are electronically based, optical implementation is particularly attractive because of its intrinsic parallelism and low energy consumption. Here, we propose and demonstrate fully-functioned all optical neural networks (AONNs), in which linear operations are programmed by spatial light modulators and Fourier lenses, and optical nonlinear activation functions are realized with electromagnetically induced transparency in laser-cooled atoms. Moreover, all the errors from different optical neurons here are independent, thus the AONN could scale up to a larger system size with final error still maintaining in a similar level of a single neuron. We confirm its capability and feasibility in machine learning by successfully classifying the order and disorder phases of a typical statistic Ising model.  The demonstrated AONN scheme can be used to construct various ANNs of different architectures with the intrinsic parallel computation at the speed of light. \\
\end{abstract}

\maketitle

Machine learning techniques, especially artificial neural networks (ANNs), have seen significant growth in the past decades and been demonstrated to be powerful or even surpass the human intelligence in various fields like image recognition, medical diagnosis and machine translation \cite{maren2014handbook, jordan2015machine}. ANNs also show great potential in scientific researches \cite{biamonte2017quantum, butler2018machine, Giuseppe2019review, sarma2019review}, especially in discovering new materials \cite{raccuglia2016machine}, classifying phases of matter \cite{carrasquilla2017machine,van2017learning}, representing variational wave functions \cite{carleo2017solving}, accelerating Monte Carlo simulations \cite{liu2017self,huang2017accelerated}, etc. They may be used to solve some problems which are intractable in conventional appraches \cite{deng2017machine, liu2017femion, zhang2017quantum, wang2017machine, deng2017quantum, peurifoy2018nanophotonic, torlai2018neural, you2018machine}. The magic power of an ANN comes from the extensive interconnections among a large amount of neurons as those in a human brain, while such a large amount of neurons and interconnections require huge computational resources (time and energy) when they are implemented with digital computers \cite{maren2014handbook}.

Unlike electrons in a digit computer, photons as non-interacting bosons could be naturally used to realize multiple interconnections and simultaneous parallel calculations at the speed of light \cite{abu1987optical, psaltis1988adaptive, caulfield1989optical,lu1990self,appeltant2011information,larger2012photonic,duport2012all}. The key ingredients of an ANN are the artificial neurons, which perform both linear and nonlinear transformations for the input signals. In most hybrid optical neural networks (ONNs), optics is mainly used for linear operations and the nonlinear functions are usually implemented electronically \cite{jutamulia1996overview, paquot2012optoelectronic, woods2012optical, hughes2018training}.  Recently ONNs basing on nanophotonic circuits \cite{shen2017deep} and light-wave linear diffraction and interference \cite{lin2018all} have been demonstrated for efficient machine learning, but the nonlinear optical activation functions are still in absence for deep networks \cite{bueno2018reinforcement,hughes2018training}. Although there have been proposals for implementing nonlinear optical activation functions \cite{george2018electrooptic,Miscuglio18}, their experimental realization has become the bottleneck for further extension of ONNs in practical applications.

\begin{figure}
\centering
\includegraphics[width=8.5cm]{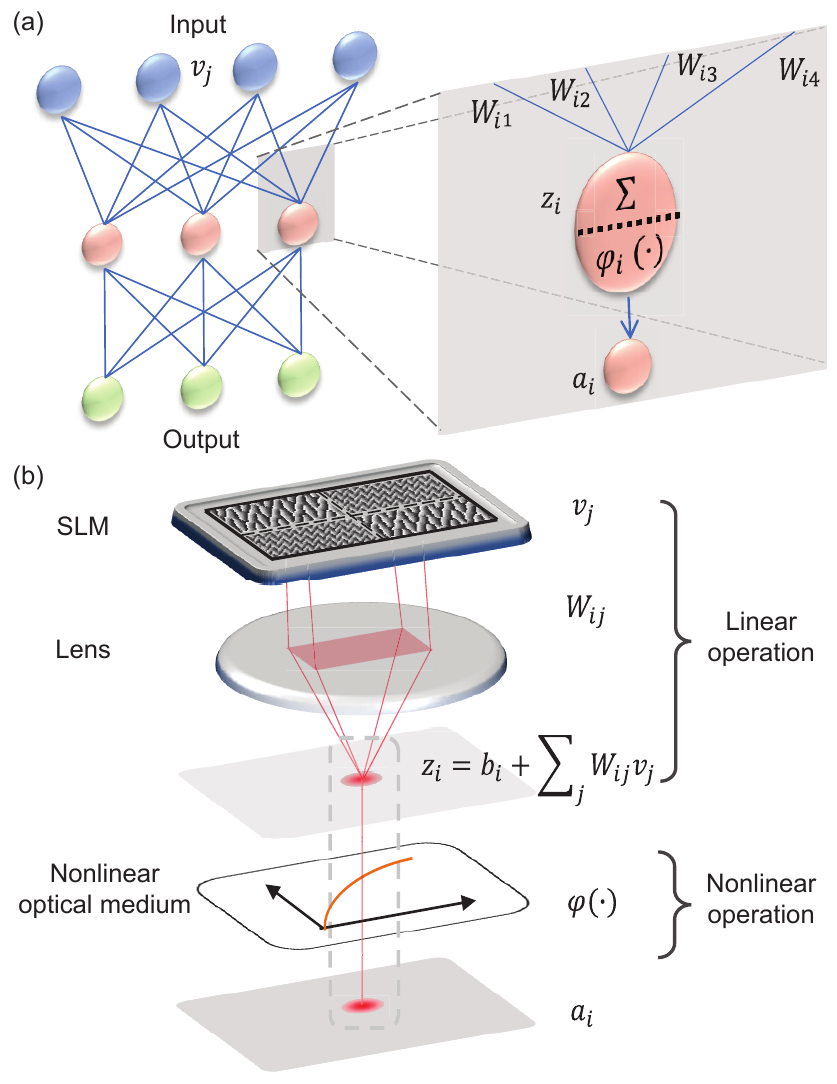}
\caption{(color online) Illustration of a general optical neural network. (a). A typical two-layer neural network. (b) Schematic of experimental realization of an optical neuron including linear and nonlinear operations. The linear operation $z_i=b_i+\sum_j W_{ij}v_j$ is achieved by combining a programable spatial light modulator (SLM) and a Fourier lens. $\varphi$ is the nonlinear activation function.} \label{fig:figure1}
\end{figure}

In this work, we demonstrate all optical neural networks (AONNs) with both tunable linear operations and non-linear activation functions in optics. We use spatial light modulators (SLMs) and Fourier lenses to implement the linear operations. The all optical non-linear activation functions are realized based on electromagnetically induced transparency (EIT) \cite{EIT_Harris, RevModPhys.77.633} - a light-induced quantum interference effect among atomic transitions. To verify the capability and feasibility of the AONN scheme, we implement several two-layer fully-connected AONNs and use them to successfully classify different phases in a prototypical Ising model.


In a typical ANN  as illustrated in Fig. \ref{fig:figure1}(a), neurons are usually arranged in layered structures without connections between different neurons in the same layer,
and the output from the neurons in one layer serves as the input for the neurons in next layer. The working principle of an artificial neuron can be abstracted into two steps: 1) receiving multiple weighted $W_{ij}$ input signals $v_j$ from neurons in the preceding layer through a linear operation and adding some linear bias $b_i$, {\emph i.e.}, $z_i=b_i+\sum_j W_{ij} v_j$, and 2) generating new output signal $a_i$ processing all the input signals through nonlinear activation functions $a_i=\phi(z_i)$. In our optical configuration, the linear operation is implemented by a SLM and an adjacent Fourier lens, and the nonlinear activitation function is realized with EIT as shown in Fig.~\ref{fig:figure1} (b). Different from the diffractive ONNs where the electric-field neurons are compelx  \cite{lin2018all}, in our AONN the signals are encoded in light power, thus $v_i, z_i, a_i \geq 0$, and the real matrix elements satisfy $1\geq W_{ij}\geq 0$.

\begin{figure}
\centering
\includegraphics[width=8.5cm]{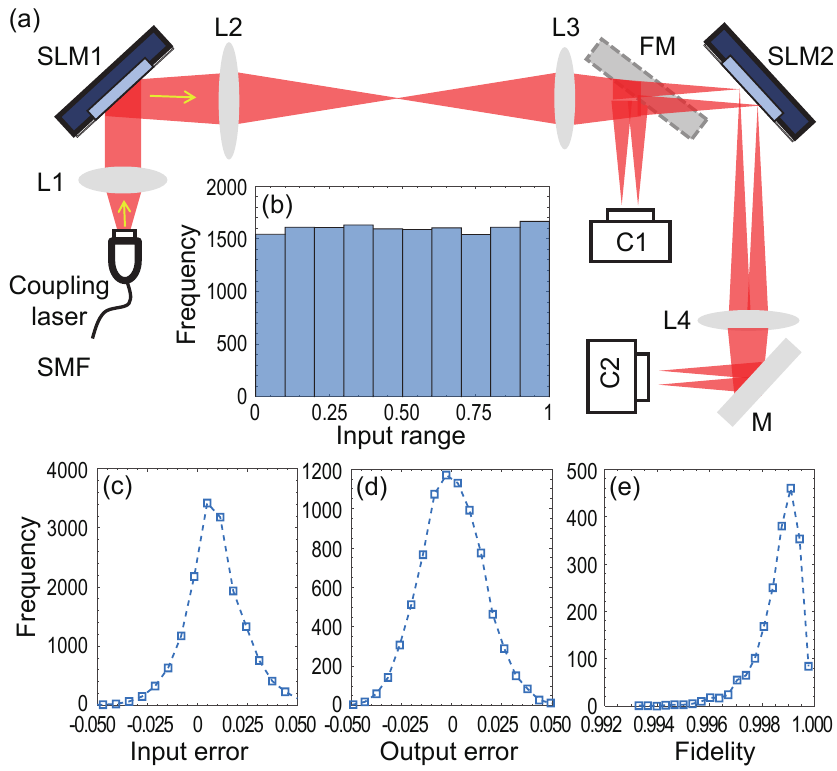}
\caption{(color online) Linear operation and characterization. (a) Optical setup.  FM: flip mirror; M: mirror; L1-L4: optical lens.  The coupling laser beam emitted from a single mode fiber (SMF) is collimated by the lens L1 (f=10 cm) and illuminates the surface of SLM1 (Leto, Holoeye), where it is selectively reflected to 8 separate light spots on the surface of SLM2 (Pluto-2, Holoeye) by a 4-f imaging system with lenses L2 (f=30 cm) and L3 (f=25 cm). A FM is inserted in the optical path to optionally reflect the beam to camera C1, which is located at the equivalent position of SLM2. SLM2 performs linear operation to tranform the 8 inputs to the 4 output beams, which are recorded by camera C2. (b) Histogram of 2000 random input vectors with elements equally distributing from 0 to 1. (c) Error distribution of the input vectors. The standard deviation (STD) is 0.012. (d) Error distribution of the output vectors for the Hankel matrix operation.  The STD is 0.014. (e) Fidelity distribution of the output vectors for the Hankel matrix operation.}
\label{fig:figure2}
\end{figure}


In the linear operation process, the incident light powers at different areas in the SLM represent the input layer nodes $v_j$. By imposing multiple phase gratings, the incident light beam $v_i$ can be split into different directions $j$ with weight $W_{ij}$. The SLM is placed at the back focal plane of the lens, which performs Fourier transform and sum all the diffracted beams in the same direction onto a spot at its front focal plane as the linear summation $z_i=\sum_j W_{ij}v_j$, as shown in Fig. \ref{fig:figure1}(b). The linear bias $b_i$ could be realized similarly. We obtain given matrix elements $W_{ij}$ following the Gerchberg-Saxton iterative feedback algorithm \cite{Gerchberg?Saxton, Yang:94}, in which high accuracy ($>95\%$) could be achieved in less than 10 iterations(See Fig. S1 in Supplementary Materials (SM) for details). Moreover, it is worth to emphasize that one great advantage of this method is that the error for a given spot does not depend on the total number of spots as long as the resolution of SLM is high enough, which is qualitatively different from the previous implementations \cite{shen2017deep,lin2018all}.

Figure \ref{fig:figure2}(a) shows the optical layout for performing the linear operation. Without losing generality, we take an 8-to-4 linear operation as an example. The coupling laser beam output from a single-mode fiber (SMF) is collimated and incident to the first SLM (SLM1), which selectively reflect 8 separate beam spots. These 8 spots are then imaged onto the second SLM (SLM2) as the input $v_j$ through a 4-f optical lens system (L2 and L3). The flip mirror (FM) and the first camera (C1) are used to monitor and measure $v_j$. The stray coupling light is blocked at the Fourier plane of lens L2.  After SLM2, each laser beam is diffracted into 4 beams. The Fourier lens L4 performs a summation operation and the 4 output spots are recorded by the second camera (C2). To characterize the accuracy of the input vectors $v_j$, we measure  error distribution of 2000 random 8-dimension input vectors with elements equally sampled from 0 to 1 [Fig. \ref{fig:figure2}(b)]. As depicted in Fig. \ref{fig:figure2}(c), we obtain very accurate input vectors with a standard deviation of only 0.017. The mean of the error is slightly off zero, due to a possible laser power drift during the measurement.

We next confirm that an arbitrary positively-valued $8\times 4$ matrix ($1\geq W_{ij}\geq 0$) can be realized by programming SLM2. We take two types of linear operations as examples. The first is the Hankel matrix $A$, a typical symmetric matrix in mathematics whose elements satisfy $A_{ij}=A_{i+k,j-k} (k=0,\dots,j-i)$ (See Eqn. S1 in SM). Because we cannot directly measure the matrix elements, we take the error distribution of output vectors using the 2000 random input vectors described previously. As shown in Fig.  \ref{fig:figure2}(d), the errors ($|\vec{z}-\vec{z'}|$) are very small with a standard deviation of 0.014, where $\vec{z}$ and $\vec{z'}$ are the exact and measured vectors respectively. Impressively, they are almost the same as the errors of input vectors even though much more operations are involved here. This result further indicates that the error could maintain in a small level even for multiple linear operations, which is crucial for large scale AONNs. For matrix production, the directions of output vectors are more useful than the exact value of different elements, the accuracy of which could be captured by the fidelity $\frac{\vec{z}\cdot\vec{z'}}{|\vec{z}||\vec{z'}|}$. As shown in Fig. \ref{fig:figure2}(e), the fidelity distribution is narrower than the error, and the mean value of fidelity is around 99.8\% for Hankel matrix. The high fidelity suggests that although there are some uncertainties for single elements, the output vectors are actually insensitive to these fluctuations. We also perform the same measurements for a random matrix and obtain similar results (see Eqn. S2 and Fig. S3 in SM for more details).

\begin{figure}
\centering

\includegraphics[width=8.5cm]{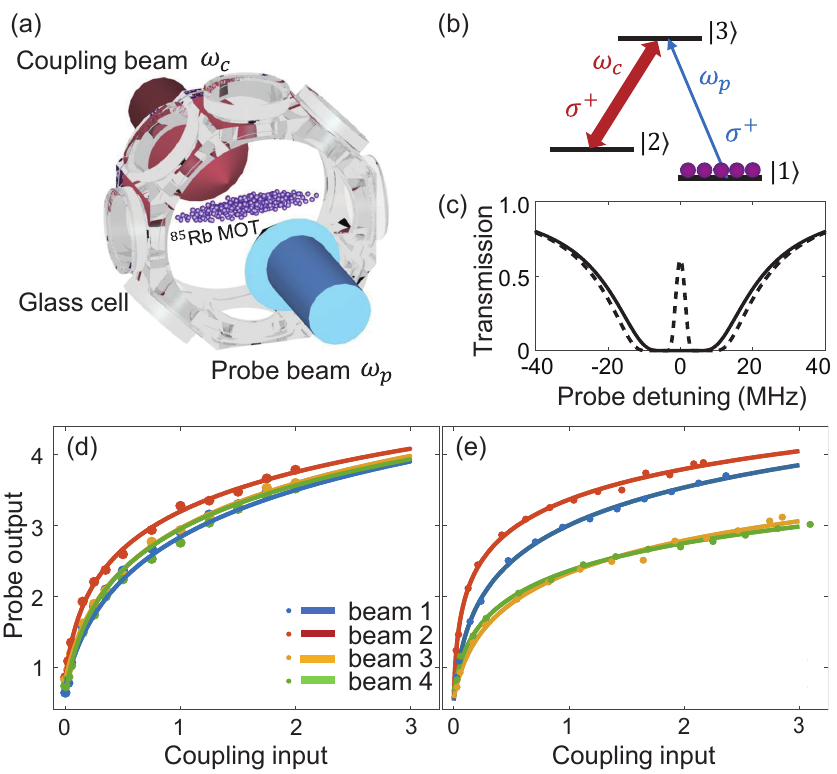}

\caption{(color online)  Realization of EIT nonlinear activation functions.  (a) EIT experimental configuration with cold $^{85}$Rb atoms in the MOT. (b) Three-level $\Lambda$-type EIT energy level diagram, where the $^{85}$Rb atomic states are $\ket{1}=\ket{5^2 S_{1/2}, F=2},\ket{2}=\ket{5^2 P_{1/2}, F=3}$ and $\ket{3}=\ket{5^2 S_{1/2}, F=3}$. Both the circularly polarized ($\sigma^+$) coupling ($\omega_c$)  laser and probe ($\omega_p$ ) laser are on resonance with the transitions $\ket{2} \to \ket{3} $ and$\ket{1} \to \ket{3} $ respectively. (c) The EIT transmission spectrum of the probe beam. The solid (dashed) line is obtained without (with) the coupling beam.  (d) and (e)  are the nonlinear transmission functions for 4 on-resonance probe beams placed at different positions of the MOT. The coupling input and probe output are scaled for fitting the neural network input-out range.} \label{fig:figure3}
\end{figure}

We now turn to the EIT nonlinear activation function. We work with laser-cooled $^{85}$Rb atoms in a dark-line two-dimensional magneto-optical trap (MOT) \cite{2DMOT} with a longitudinal length of 1.5 cm and  an aspect ratio of 25:1, as shown in Fig. \ref{fig:figure3}(a). The atoms are prepared in the ground state $|1\rangle$, as shown in the atomic energy level digram in Fig. \ref{fig:figure3}(b). The circularly polarized ($\sigma^+$) coupling laser ($\omega_c$) beams, which are from the outputs of the linear operation, are on resonance to the atomic transmission $|2\rangle\leftrightarrow|3\rangle$ and incident to the atomic cloud along its transverse direction.  A counter-propagating probe laser beam ($\omega_p, \sigma^+$) is on resonance to $|1\rangle\rightarrow|3\rangle$. In absence of the coupling beam, the atomic medium is opaque to the resonant probe beam which is maximumly absorbed by the atoms, as shown as the solid curve in the transmission spectrum of Fig. \ref{fig:figure3}(c). While in presence of the coupling beam, the quantum interference between the transition paths leads to an EIT \cite{EIT_Harris, RevModPhys.77.633} transparency spectral window as shown as the dashed curve  in Fig. \ref{fig:figure3}(c), where the on-resonance peak transmission and the bandwidth are controlled by the coupling laser intensity. The on-resonance probe laser beam output can be expressed as
\begin{equation}
I_\text{p,out}=I_\text{p,in}e^{-OD\frac{4\gamma_{12}\gamma_{13}}{\Omega_c^2+4\gamma_{12}\gamma_{13}}}=\varphi(\Omega_c^2),
\label{eq:EIT}
\end{equation}
where $I_\text{p,in}$ and $I_\text{p,out}$ are the input and output probe beam intensity, $OD$ is the atomic optical depth on the $|1\rangle\rightarrow|3\rangle$ transition, and $\gamma_{ij}$ is the dephasing rate between the states $\ket{i}$ and $\ket{j}$. For $^{85}$Rb atoms, $\gamma_{13}=2\pi \times 3$ MHz, and the non-zero ground-state dephasing rate $\gamma_{12}$ can be tuned by the stray background magnetic field. $\Omega_c$ is the coupling field Rabi frequency and its square is proportional to the coupling laser intensity ($\Omega_c^2\propto I_c$). As shown in Eq. (\ref{eq:EIT}), the probe beam intensity is nonlinearly controlled by the coupling beam intensity. The nonlinear activation function $\varphi$ is achieved by taking the coupling intensity as the input and the transmitted probe intensity as the output. In experiment, the input probe beam is collimated and its beam size is large enough to cover the entire coupling beam profile. Moreover, Equation (\ref{eq:EIT}) also indicates that the the nonlinear activation function is determined by $OD$ and $\gamma_{12}$ whose values are different at different positions of the MOT. Therefore, by placing the counter-propagating coupling-probe beams at different positions of the MOT, we can achieve different nonlinear activation functions for different neurons. Figure \ref{fig:figure3}(d) shows nearly identical nonlinear activation functions are obtained by carefully positioning the four input coupling beams. We can also assign these 4 neurons with different nonlinear activation functions, as shown in Fig. \ref{fig:figure3}(e). Clearly, the errors from different nonlinear active functions are also independent. Together with the same advantages of linear operations realized by SLMs and lenses, the AONN scheme could be expected to scale up to large size with error maintained in a small level.

\begin{figure*}
\centering
\includegraphics[width=17 cm]{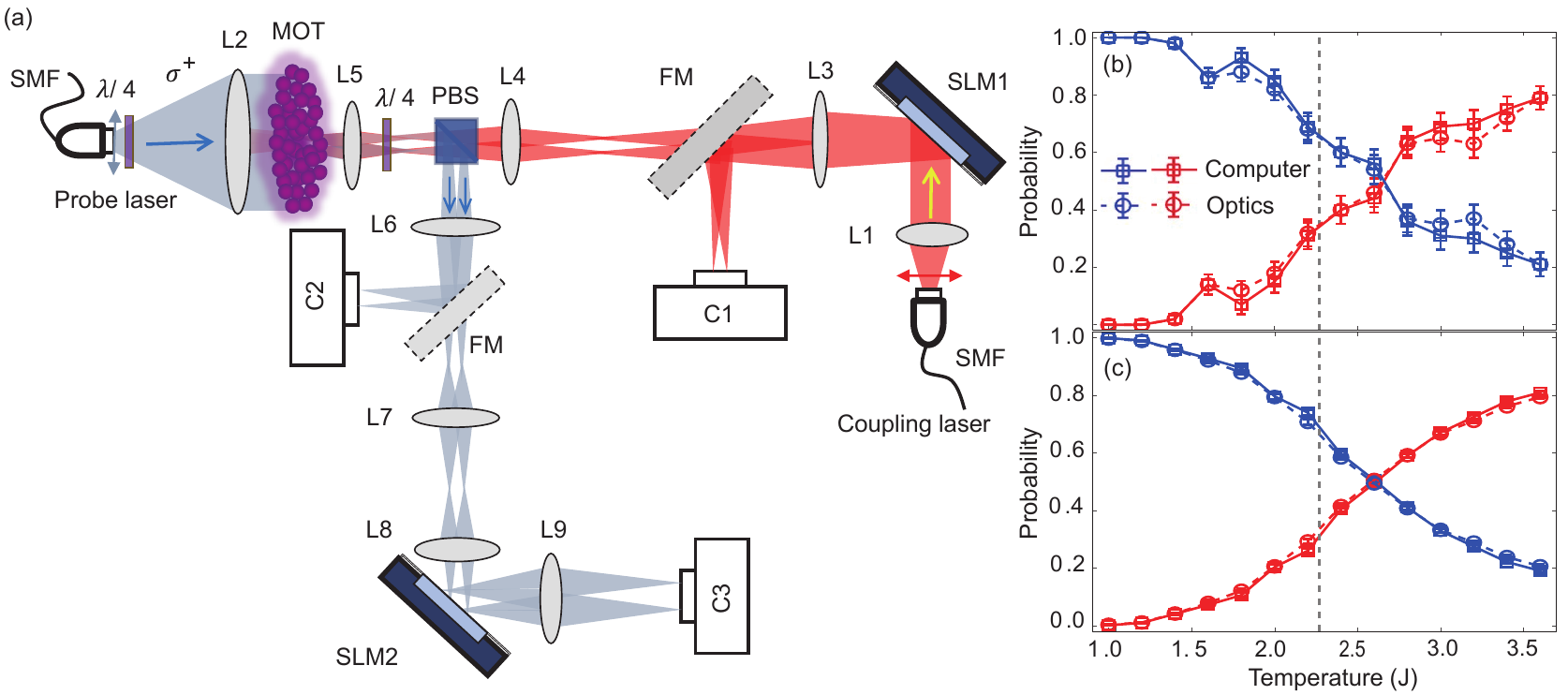}
\caption{(color online) Fully functioned 2-layer all optical neural network (AONN). (a) Experimental configuration of the two-layer AONN. The input layer is the pattern encoded on SLM1 whose area is divided into multiple subareas. The first layer consists of a linear operation done by SLM1 and EIT nonlinear activation functions at MOT. The second layer contains SLM2 which converts four beams into two output beams at camera C3. The collimated coupling laser beam passing lens L1 (f=10 cm) is incident on the SLM1 which generate four beams at the focal plane of L3 (f=75 cm). A flip mirror (FM) and camera C1 are used to monitor this linear operation. The four beams are imaged on the MOT through a 4-f system consisting of L4 (f=75 cm) and L5 (f=5 cm).  A collimated probe laser is propagating along the opposite direction of the coupling beam, which is imaged on camera C2 through L5 and L6 (f=45 cm). With further amplification by L7 (f=7.5 cm) and L8 (f=45 cm), four beams are incident on SLM2 and generate two beams and then are focused on camera C3. (b) and (c) are the mean probability of the order (blue) and disorder (red) phases as functions of temperature for 100 and 4000 configurations respectively. The comparison is made between the results from the AONN (circle) and the computer NN (square). } \label{fig:figure4}
\end{figure*}

After the demonstration of linear and nonlinear operations, we are ready to assemble a fully functioned AONN using SLMs, lenses, MOT, the coupling and probe laser beams. Next we will show that we can actually apply such an AONN to classify different phases in condensed matter physics. It has been demonstrated recently that neural networks have great potentials to identify different phases including both symmetry-breaking phases \cite{carrasquilla2017machine,van2017learning} and topological phases \cite{carrasquilla2017machine,zhang2017quantum}. Here we take the prototypical two-dimension Ising model on square lattice as an example for the demonstration. The Ising model could be written as
$H(\sigma)=-J\sum_{\left<ij\right>}\sigma_i \sigma_j$,
where $\sigma_i=\pm1$ represents localized spin on site $i$, and $\left<ij\right>$ takes the summation over all the nearest neighbors. It is well known that there will be a continuous phase transition with the critical temperature as $T_c= \frac{2}{\ln{(1+\sqrt{2})}}|J|$. In the following simulations and experiments, the interaction strength is set to be $J=1$, and we take lattice size $L=4$ with periodic boundary conditions as examples.
We found a 2-layer fully-connected neural network (FCNN) could work well by using the experimentally measured nonlinear activation functions in Fig. \ref{fig:figure3}(e). Same as that in a conventional electronic computer, our 2-layer AONN consists of one input layer, one hidden layer and one output layer. Optical signal propagates from one layer to the next layer through optical operation units. At the hidden layer, optical information is processed by optical nonlinear active functions before propagating to the output layer.

The detailed optical implementation of the 2-layer AONN is illustrated in Fig. \ref{fig:figure4}(a). The input layer contains $L^2$ neurons with $L$ as the linear system size. For the hidden and output layers, there are 4 and 2 neurons, respectively.  In this particular configuration, because the input values are binary (0 or 1), the coupling beam input vector and the first linear operation can be realized by a single SLM, as shown as SLM1 in Fig. \ref{fig:figure4}(a) (See Eqn. S5 in SM for details).  The horizontally polarized output coupling beam passes through a polarizing beam splitter (PBS) and is incident to the cold atoms in MOT. The 4 transmitted counter-propagating and vertically polarized probe beams are reflected by the PBS and enters SLM2 for a second linear operation which reduce the 4 inputs to 2 outputs recorded by camera C3. The FMs and cameras C1 and C2 are used for configuring the network parameters.

The 2-layer FCNN is initially trained in computer and later finely tuned in the AONN through supervised learning from the labelled raw configurations generated by Monte Carlo (MC) simulations. By using the configurations generated at lower and high temperatures, the AONN learns to label them as the ordered or disordered. The fraction of ordered or disordered configurations among all the samples can be regarded as the probability being in order or disorder states. Then the crossing temperature with both probabilities at $50\%$ could be taken as the phase transition point. Thus, after machine learning, the 2-layer AONN is applied to identify different phases from configurations sampled by MC simulation at intermediate temperature and find the critical phase transition temperature (See Supplementary training of two-layer AONNs for details).

Figures \ref{fig:figure4}(b) and (c) plot the mean probability of configuration sets generated at different temperatures for $L^2=4^2=16$ inputs. In general, the experimental results can well reproduce the whole phase diagram even though we only train the AONN at the temperatures far away from the critical temperature. The experimental phase transition temperature is close to the analytical thermodynamical limit represented by the vertical dash line as the number of sites goes to infinite. It suggests that the AONN successfully capture the essential features, which distinguish the order and disorder phases. To clearly show the performance of our AONN, we first intentionally did the experiments by using only 100 configurations and the results are shown in Fig. \ref{fig:figure4}(b). In the sense of phase classification, the results are not very good. It is reasonable since we only use very few configurations in a very large configure space (100 out of $2^{4\times4}$ $\sim 0.15\%$). However, the results from our AONN and the computer FCNN for the same configurations are nearly identical for all the temperatures, which clearly shows that our AONN has the same accuracy as the well trained computer-based NN. To further demonstrate the capability of our AONN, we then did the experiments with 4000 configurations. As expected, the phase transition curves become smoother because the random errors from statistical fluctuations are strongly reduced, and the optical results are almostly the same as the computer data as shown in Fig. \ref{fig:figure4}(c). All these results confirm that our implementation of AONN is successful and indeed capable to classify different phases for the Ising model.
In summary, we demonstrate an AONN scheme with both tunable linear operations and nonlinear activation functions.  The linear interconnections are realized using SLMs and optical lenses. The EIT optical nonlinear activation functions are based on quantum interference.  For demonstration, we constructed a 2-layer FCNN for classifying the phases of a prototypical Ising model. While in this work, we focus mainly on the feasibility of the linear operations and nonlinear activation functions, which are the key ingredients of an ANN, the AONN is scalable to a larger system size with more SLMs and EIT nonlinear media, with a cost of longer learning time in finely tuning the network parameters.
The reasons are two-fold: (1) As the magic power of ANNs comes from the extensive interconnections between a large mount of neurons, ANNs are error-tolerant and robust against small local random errors, which means even the local parameters are not precise we can still get very good results as long as the number of neurons is large enough just as human brains. For most of problems, more neurons in ANNs usually give better performance; (2) As clearly demonstrated in our experiments, the final error of our AONN is insensitive to the total number of neurons and the error could maintain in a similar level as a single neuron even for large scale AONNs. Such a big advantage inherits from the fact that all the linear and nonlinear active functions in our AONNs are independent and the errors from different optical neurons will not accumulate but may cancel with each other. Implementing a large scale AONN requires more engineering resources and we leave it for future work.


This work was supported by the Hong Kong Research Grants Council (Projects Nos. C6005-17G and ECS26302118). B.L. and Y.C.C. acknowledge the support from the Undergraduate Research Opportunities Program at the Hong Kong University of Science and Technology.

\bibliography{AONN}


\end{document}